\documentclass[article,twocolumn,twocolappendix]{aastex6}

  \usepackage{amsmath}
  \usepackage{tabularx}
  \usepackage{color}
  \usepackage{graphicx}
  \usepackage{rotating}

\begin{document}


\title{Spectroscopy of 10 $\gamma$-ray BL Lac objects at high redshift  \\
     }



\author{Simona Paiano\altaffilmark{1,2,3},  Marco Landoni\altaffilmark{3}, Renato Falomo\altaffilmark{1}, Aldo Treves\altaffilmark{4}, Riccardo Scarpa\altaffilmark{5,6}}




\altaffiltext{1}{INAF, Osservatorio Astronomico di Padova, Vicolo dell'Osservatorio 5 I-35122 Padova (PD) - ITALY}
\altaffiltext{2}{Universit\`a di Padova and INFN, Via Marzolo 8, I-35131 Padova (PD) - ITALY}
\altaffiltext{3}{INAF, Osservatorio Astronomico di Brera, Via E. Bianchi 46 I-23807 Merate (LC) - ITALY}
\altaffiltext{4}{Universit\`a degli Studi dell'Insubria, Via Valleggio 11 I-22100 Como (CO) - ITALY}
\altaffiltext{5}{Instituto de Astrofisica de Canarias, C/O Via Lactea, s/n E38205 - La Laguna (Tenerife) - SPAIN}
\altaffiltext{6}{Universidad de La Laguna, Dpto. Astrofísica, s/n E-38206 La Laguna (Tenerife) -  SPAIN}

\begin{abstract}

We present high S/N optical spectra of 10 BL Lac objects detected at GeV energies by \textit{Fermi} satellite (3FGL catalog), for which previous observations suggested that they are at relatively high redshift.  
The new observations, obtained at the 10 m Gran Telescopio Canarias, allowed us to find
the redshift for J0814.5+2943 (z = 0.703) and we can set spectroscopic lower limit for J0008.0+4713 (z~$>$~1.659) and  J1107.7+0222 (z~$>$~1.0735) on the basis of Mg II intervening absorption features.
In addition we confirm the redshifts for J0505.5+0416 (z~$=$~0.423) and for J1450+5200 (z~$>$~2.470). 
Finally we contradict the previous $z$ estimates for five objects (J0049.7+0237, J0243.5+7119, J0802.0+1005, J1109.4+2411, and J2116.1+3339).

\end{abstract}


\keywords{BL Lac object spectroscopy ---  Redshift }

\section{Introduction} \label{sec:intro}

Blazars are active galactic nuclei (AGN) where the relativistic jet is pointing in the observer's direction. 
They are characterized by high variability in all bands, and large polarization. 
The spectral energy distribution (SED) exhibits two broad bumps, one in the IR-X-ray band, and one in the MeV-TeV band. The former is due to synchrotron radiation produced by the relativistic electrons in the jet, while the latter in most models is due to the Compton scattering of the same electrons \citep[e.g.][]{maraschi1992, dermer1993, ghisellini2009a}. 
 In some cases the thermal contribution due to the AGN accretion disk is also visible \citep[see e.g.][for a recent review]{madejski2016}.

Blazars are usually divided in two classes, BL Lac Objects (BLLs) and Flat Spectrum Radio Quasars (FSRQs), depending on the strength of the broad emission lines with respect to the continuum.  
A more physical distinction refers to the comparison between the  broad line region luminosity and the Eddington luminosity. 
FSRQs have radially efficient accretion disk, while BLLs are not able to photoionize gas in the clouds of the broad line region, explaining the laking of these features in the majority of their spectra \citep[e.g.][and references therein]{ghisellini2017}. 
Note that this classification requires the knowledge of the mass of the accreting black hole, and of the distance, which for broad emission line AGNs can be easily determined by spectroscopy.  
However, this becomes arduous for the BLLs due to the weakness of the spectral lines.

The advent of the \textit{Fermi} gamma-ray observatory (1-100 GeV), starting observations in 2008 \citep{atwood2009}, with its systematic scanning of the entire sky every 3 hours, has substantially modified the study of blazars, which previously was based mostly on radio and X-ray surveys. 
In fact it was shown that blazars dominate the extragalactic gamma-ray sky \citep{3FGL}.
The third AGN FERMI/LAT catalogue \citep[3LAC,][]{3lac} contains 1738  blazars, compared with the $\sim$~3000 $\gamma$-ray detected sources, where 662 are classified as BLL and 491 as FSRQ. The remaining blazars are reported as of uncertain type.

It is worth to note that for a large fraction of the BLLs the
redshift is still unknown or highly uncertain. Based on the
present statistics it was nevertheless proposed that on average
BLLs have lower redshift and smaller high energy (HE; $>$~20MeV) $\gamma$-ray luminosity
than FSRQs \citep{ghisellini2017}.  This proposal, however,
could be biased since at high redshift the number of robustly
detected BLLs is drastically reduced due the difficulty of measuring the redshift \citep[e.g. ][and references therein]{falomo2014}. 
Moreover the uncertainty of the redshift hampers to perform a sound comparison of the characteristics of the multiwavelength SED between the two classes of blazars for which both the bolometric luminosity and redshift are needed \citep[see the so-called \textit{blazar sequence},][]{fossati1998}.

The determination of the redshift of BLLs is 
also important to
characterize the properties of the extra
galactic background light \citep[EBL, e.g. ][and references therein]{franceschini2008}. 
The Very High Energy sky (VHE; $>$~100 GeV), observed with Cherenkov telescopes, 
is mainly dominated by BLLs (in the TeVcat\footnote{http://tevcat.uchicago.edu/}
there are  60 BLLs against 6 FSRQs). Their
energetic $\gamma$-rays  can interact with lR-optical EBL
photons to produce e$^{-}$/e$^{+}$ pairs, resulting in a clearly
detectable absorption in the GeV-TeV BLL spectrum  starting at frequencies and with optical depth
that depend on the redshift of the $\gamma$-ray source and is more pronounced in the 0.5~$<$~z~$<$~2
interval.
At higher $z$ the absorption due to pair production moves to \textit{Fermi} energies, completely extinguishing the source in the VHE regime.
Although a significant number of FSRQ detections, up to $z$~$>$~4 already exist \citep{ackermann2017}, at the TeV energies, due to their Compton inverse peak position, only a small fraction of them are detected. Therefore the identification of high redshift\footnote{Few known redshifts of BLLs range between 0.5 and 2.5.} BLLs at these energies is particularly challenging in order to study the earliest EBL components due to the first-light sources (Population III stars, galaxies or quasars) in the universe \citep{franceschini2017}. 

In the framework of our long-term optical spectroscopy program at large (8 - 10 m) telescopes, aimed at determining the redshift of the BLLs \citep{sbarufatti2005, sbarufatti2006,  sandrinelli2013, landoni2012, landoni2013, landoni2014, landoni2015, paiano2016, paiano2017}, we concentrate here on 10 BLLs detected by Fermi satellite 
with unknown or very uncertain redshift.

In this work we assume the following cosmological parameters: H$_0=$ 70 km s$^{-1}$ Mpc$^{-1}$, $\Omega_{\Lambda}$=0.7, and $\Omega_{m}$=0.3.

\section{Sample, reduction and data analysis} \label{sec:sample}
 
We searched for BLLs  that are candidates for being at high redshift (z~$>$~1) in the \textit{Fermi} 3LAC catalog.  
These objects have uncertain redshift and, in most cases, conflicting values are reported in the literature mainly due to low S/N spectra. 
Considering only the sources that are well observable from La Palma site the selection produced 18 targets and we obtain observations for 10 of them (see Table \ref{tab:table1}). 

The observations were gathered in Service Mode at the GTC using the low resolution spectrograph OSIRIS \citep{cepa2003}. 
The instrument was configured with the grisms R1000B and R1000R\footnote{http://www.gtc.iac.es/instruments/osiris/osiris.php}, in order to cover the whole spectral range 4100-10000~$\textrm{\AA}$, and with a slit width~$=$~1'' yielding a spectral resolution $\lambda$/$\Delta\lambda$~$=$~800.

For each grism, we obtained three individual exposures (with exposure time ranging from 150 to 1200 seconds each, depending on the source magnitude), which were combined into a single average image, in order to perform an optimal cleaning of cosmic rays and CCD cosmetic defects.
Wavelength calibration was performed using the spectra of Hg, Ar, Ne, and Xe lamps and providing an accuracy of 0.1~$\textrm{\AA}$ over the whole spectral range.  
For each object the spectra obtained with the two grisms were merged into a final spectrum covering the whole desired spectral range.
Spectra were corrected for atmospheric extinction using the mean La Palma site extinction table\footnote{https://www.ing.iac.es/Astronomy/observing/manuals/}. 
Relative flux calibration was provided by spectro-photometric standard stars secured during the same nights of the target exposure.  
The observation strategy and the data reduction followed the same procedure reported in the \citet{paiano2017} and detailed information on the observations are given in Table \ref{tab:table2}.

\section{Results} \label{sec:results}

The optical spectra of the targets are presented in Fig. \ref{fig:spectra}.
In order to emphasize weak emission and/or absorption features, we show also the normalized spectrum. 
This was obtained by dividing the observed calibrated spectrum by a power law continuum fit of the spectrum, 
excluding the telluric absorption bands (see Tab. \ref{tab:table3}).
The normalized spectra were used to evaluate the signal-to-noise ratio (S/N) in a number of different spectral regions.  
On average, the S/N ranges from 10 to 200 depending on the wavelength and the magnitude of the source (Tab. \ref{tab:table3}).
These spectra can be accessed at the website http://www.oapd.inaf.it/zbllac/.

All spectra were carefully inspected to find emission and absorption features.  When a possible feature was identified, we determined its reliability checking that it was present in the three individual exposures (see Sec. \ref{sec:sample} for details).  

We were able to detect stellar spectral features of Ca~II (3934,3968) for 3FGL~J0505.5+0416 and 3FGL~J0814.5+2943.
For two sources, 3FGL~J0008.0+4713 and 3FGL~J1107.5+0222, we detect strong intervening absorption system due to Mg~II~(2800) , allowing to set a spectroscopic lower limit of their redshift. 
Finally in one object, 3FGL~1450.9+5200, we detect intervening absorption systems due to C~IV~(1548) and Ly$_{\alpha}$~(1216) at two different redshifts. 
For five other objects the spectrum appears featureless in contrast with the previous claimed redshift values.
Details are reported in Fig. \ref{fig:spectraCU} and in Tab. \ref{tab:line}.

Starting from the basic assumption that all BLLs are hosted by a massive elliptical galaxy, one si able to detect faint absorption features from the starlight provided that the S/N and the spectral resolution are sufficiently high. 
According to the scheme outlined in \citet{paiano2017}, in the case of no detection of spectral features it is also possible to set a lower limit to the redshift based on the minimum Equivalent Width (EW) spectrum \citep[see Appendix A of ][for details]{paiano2017}.
The results about the redshift lower limits obtained for the whole sample are summarized in Table \ref{tab:table3} and details about the optical spectra and redshift estimates for each individual object of our sample are given in Sec. \ref{individual}.

\section{Notes for individual sources } \label{sec:notes}
\label{individual}

\begin{itemize}
\item[] \textbf{3FGL J0008.0+4713}:
From an optical spectrum provided by \citet{kock1996}, a redshift of 0.28 was proposed, from absorption features of its host galaxy (the spectrum is not published).
However, the object appears unresolved from optical images \citep{nilsson2003}.
A more recent and rather noisy spectrum from \citet{shaw2013} gives a tentative high redshift z~$=$~2.1, based on the onset of the Lyman-$\alpha$ forest.
In our spectrum we clearly detect an intervening  absorption doublet at 7440 $\textrm{\AA}$ (see Table 4) that we identify with Mg II (2800) at z~$=$~1.659. 
No other emission or absorption features are found.
The continuum is well fitted by a power law with the rather flat spectral index ($\alpha=$+0.74, $ F_{\nu} \propto \nu^{\alpha}$), suggesting dust extinction possibly associated to intervening gas at z~$=$~1.659.
This target is therefore one of the highest redshift BLL known thus far.

\item[] \textbf{3FGL J0049.7+0237}:
First optical spectrum was secured by \citet{dunlop1989} finding the source featureless. 
Same result was obtained few years later by \citet{allington1991}.
A superior quality optical spectrum was then published by \citet{sbarufatti2006}, that confirm the featureless spectrum of the source, while \citet{shaw2013} claim the detection of the broad emission at $\sim$~6930$\textrm{\AA}$ identified as Mg II (2800), suggesting z~$=$~1.474.
From our spectrum (S/N~$\sim$~100), we contradict the presence of the above feature, therefore the redshift is still unknown.
We set a lower limit based on the non-detection of the starlight (see Sec. \ref{sec:results}) of z~$>$~0.55.

\item[] \textbf{3FGL J0243.5+7119 }:
A featureless optical spectrum was reported by \citet{stickel1996}.
A possible intervening absorption of Mg II (2800) at 5595~$\textrm{\AA}$ is claimed by \citet{shaw2013} yielding a redshift limit of z~$>$~1.
We do not confirm this absorption doublet in our spectrum and note that at this wavelength there is a very strong night sky emission. 
We also note that this source is at low galactic latitude (b=10$^{\circ}$) and therefore the source is severely absorbed ( E(B-V)~=~0.7 ). 
We set a lower limit of z~$>$~0.45  based on lack of starlight features.

\item[] \textbf{3FGL J0505.5+0416}:
This is a radio source \citep{bennett1986, bauer2000} well-detected at X-ray and gamma-ray frequencies \citep{voges1999, acero2015}. 
The first optical spectrum is reported in \citet{laurent1998} that fails to detect any lines.
Optical images, obtained by \citet{nilsson2003}, were able to resolve the host galaxy, indicating a relatively low redshift. 
The optical spectrum obtained by \citet{pita2014} shows absorption features of the host galaxy yielding a redshift of 0.424. 
We confirm the Ca II doublet (3934, 3968), which in the Pita et al. spectrum was in the merging region of the UVB and VIS arm of the instrument, and we also detect the absorption line due to the G-band (4305) at z~$=$~0.423 (see Fig. \ref{fig:spectraCU}).

\item[] \textbf{3FGL J0802.0+1005}:
The object was identified as a BLL by \citet{plotkin2010} on the basis of the SDSS spectrum , and it is not yet detected in the X-ray band and the counterpart is very faint in the radio regime \citep{nvss1998}.
It was observed twice by SDSS and two different redshifts were proposed from the automatic line identification (z~$=$~0.06 and z~$=$~0.842).
From our visual inspection of these two spectra, no significant presence of emission or absorption lines are seen. 
Our optical spectrum confirms the featureless continuum described by a power law of $\alpha~=$~-0.77 and we can set a relatively high redshift limit of z $>$~0.58.

\item[] \textbf{3FGL J0814.5+2943}:
\citet{white2000} found a featureless optical spectrum. 
In our high S/N$\sim$160, we detect an absorption doublet system at 6699-6759 $\textrm{\AA}$ that we identify as Ca II at z~$=$~0.703 from the starlight of the host galaxy (see Fig. \ref{fig:spectraCU}). 
We note that the redshift value reported by NED (z~$=$~1.083) based on the SDSS spectrum is contradicted by our spectrum.

\item[] \textbf{3FGL J1107.5+0222}:
In \citet{plotkin2010}, the authors found a featureless optical spectrum based on SDSS data. 
We clearly detect an absorption doublet at 5797,5812 $\textrm{\AA}$ that we identify with Mg II (2800) intervening system at z~$=$~1.0735 (see Fig. \ref{fig:spectraCU}).

\item[] \textbf{3FGL J1109.4+2411}:
The source was discovered and identified as a BLL from X-ray Einstein Sky Survey \citep{perlman1996}. 
The host galaxy was detected from images by HST snapshots suggesting a redshift z$\sim$0.5 \citep{sbarufatti2005, falomo1999, sbarufatti2005}.
The optical spectrum obtained by SDSS is found featureless\footnote{the SDSS automatic line identification gives z=1.22} \citep{shaw2013} and also an ESO VLT spectrum reported by \citet{landoni2013} appears featureless. 
We obtain an optical spectrum with a S/N ranging from 20 to 70, and no significant emission or absorption lines are detected with a minimum EW of $\sim$~0.35 $\textrm{\AA}$ that allows to set a redshift lower limit of $>$~0.5.

\item[] \textbf{3FGL J1450.9+5200}:
The source was proposed as a very high redshift BLL by \citet{plotkin2010} on the basis of intervening absorption of C~IV(1548) and Ly$_{\alpha}$ (1216) in the SDSS spectrum at redshift of z~$=$~2.474.
We also detect the presence of the absorption line at 5372 $\textrm{\AA}$ identified as C~IV (1548) intervening gas, and at the same redshift we see a strong Ly$_{\alpha}$ (1216) absorption.
In addition, we find an absorption feature at 5127 $\textrm{\AA}$ that we interpret as a second C~IV (1548) intervening system at z~$=$~2.312.
Finally other absorption lines due to Ly$_{\alpha}$ are observed at $\lambda<4219~\textrm{\AA}$. The redshift of this source is thus z~$>$~2.474.

The source is detected by \textit{Fermi} up to the 60-80 GeV interval \citep{3fhlcatalog}. 
The optical depth at z~$=$~2.470 for pair production in the EBL is $\tau \sim 1$ (see Fig. 10 of  Franceschini et al. 2017, in press).

\item[] \textbf{3FGL J2116.1+3339}:
The source was originally detected in the radio-band (B 2114+33) and it appears in ROMA BZCAT \citep[BZB J2116+3339][]{massaro2015},  which reports a radio, optical (R=15.4), X-ray, and gamma-ray flux. 
An optical spectrum is given by \citet{shaw2013} which to us appears featureless, although the authors claim a redshift of 1.596 based on a very dubious extremely faint C~IV (1548) emission.
In contrast in our spectrum we do not detect any emission or absorption lines with EW~$>$~0.30 $\textrm{\AA}$ and we set a redshift lower limit of $>$~0.25 .

\end{itemize}

\newpage 
\section{Conclusion} \label{sec:discu}

The discovery of hundreds of BLLs by the \textit{Fermi} satellite motivates accurate optical spectroscopic studies, which in most cases require the use of large telescopes. 
In fact, the information on the redshift derives from the detection of very weak absorption or emission lines, at the source or in the intervening material. 
Our paper has focused on 10 sources detected at GeV energies, for which we present spectra of very good quality with S/N values ranging up to $\sim$200. 
They were chosen among objects already observed with smaller telescopes or with a poor S/N spectrum and there was some indication that they were at relatively high z.
For five out of 10 sources studied, spectral features were detected, while for the rest the only information on the redshift is a lower limit based on the absence of absorption lines from the host galaxy. 
For 8 out of the 10 sources the redshifts are above 0.5, and two of them are two of the farthest BLLs known till now,  3FGL J0008.0+4713 (z~$>$~1.659) and 3FGL J1450.9+5200 (z~$>$~2.470).
In spite of the improvement of the observing facilities (large telescope and modern instrumentation), the redshift determination of some high redshift BLLs remains rather arduous. 
However its knowledge is crucial for the advancing of our understanding of BLLs in a cosmic context.

\newpage

\begin{table*}
\caption{THE SAMPLE OF 3FGL/LAC BL LAC OBJECTS}\label{tab:table1}
\centering
\begin{tabular}{llccccl}
\hline 
3LAC name &  Other name  &  RA   & $\delta$   &     V       & $E(B-V)$     &     tentative redshifts\\
                     &                       & (J2000)     & (J2000)    &            &  $(degree)$   &                       \\
\hline 
3FGL J0008.0+4713      &   BZB J0007+4712    &   00:08:00.0    &   +47:12:08    &   18.30  &   0.08    &   0.28 , 2.10      \\
3FGL J0049.7+0237      &   PKS 0047+023        &   00:49:43.2   &    +02:37:04    &   18.00  &  0.01     &  1.44 , 1.474    \\
3FGL J0243.5+7119      &   BZB J0243+7120    &   02:43:30.9    &    +71:20:18    &   19.20  &   0.70   &   $>$0.998,  ?      \\
3FGL J0505.5+0416      &   BZB J0505+0415	&   05:05:34.8    &    +04:15:55    &   16.70  &  0.07  &  0.424 , ?          \\             
3FGL J0802.0+1005     &    BZB J0802+1006    &   08:02:04.8   &    +10:06:37    &    17.42  &  0.02    & 0.842,  ?          \\ 
3FGL J0814.5+2943      &   EXO 0811+2949     &   08:14:21.3    &    +29:40:21    &   18.80  &  0.03    &  1.084              \\ 
3FGL J1107.5+0222      &   BZB J1107+0222	&   11:07:35.9    &    +02:22:25    &   18.97  &  0.03    & ?                      \\    
3FGL J1109.4+2411      &   1ES 1106+244	&   11:09:16.1    &    +24:11:20    &   18.70  &  0.02    &  0.482,  1.221   \\
3FGLJ1450.9+ 5200      &   BZB J1450+5201    &    14:50:59.9  &    +52:01:11    &    18.90  &  0.02   & 2.471,   2.474    \\        
3FGL J2116.1+3339      &    2FGL J2116+3339	&   21:16:14.5    &   +33:39:20     &   16.30  &  0.10   &  0.35 , 1.596       \\
\hline
\end{tabular}
\tablenotetext{}{
\raggedright
\footnotesize \texttt{Col.1}: 3FGL/LAC name of the target; \texttt{Col.2}: Other name of the target; {Col.3}: Right Ascension; \texttt{Col.4}: Declination; \texttt{Col.5}: V-band magnitudes taken from NED; \texttt{Col.6}: $E(B-V)$ taken from the NASA/IPAC Infrared Science Archive (https://irsa.ipac.caltech.edu/applications/DUST/); \texttt{Col.7}: Tentative redshift  taken from the literature. For references see \citet{plotkin2010, shaw2013}, 3LAC catalog \citep{3lac}, and NED.}
\tablenotetext{}{
\raggedright
 } 
\end{table*}

\newpage

\begin{table*}
\caption{LOG OF GTC OBSERVATIONS OF 3FGL/LAC BL LAC OBJECTS}
\label{tab:table2}
\centering
\begin{tabular}{c|cccc|cccc}
\hline
 & \multicolumn{4}{c|}{Grism B}    &  \multicolumn{4}{c|}{Grism R}   \\
\hline
Obejct          &   t$_{Exp}$  (s)    &       Date            & Seeing  ('')   &   r  &    t$_{Exp}$   (s)    &         Date         & Seeing  ('')  & r  \\
                    &                             &                          &                      &       &                               &                        &                     & \\
\hline
3FGL J0008.0+4713    &	1800    &	2015 Oct 04      &   1.5     & 18.4 &      	1800	 &	2015 Nov 23	& 2.0  &	18.5	\\
3FGL J0049.7+0237    &	3600    &	2015 Nov 23     &   1.5     &  18.5 &     	3600	 &	2015 Nov 23	& 1.9   &	18.5  \\
3FGL J0243.5+7119    &	3600    &	2015 Nov 20     &   1.6     &  18.9 &    	2700	 &	2015 Nov 20	&  2.2  &	18.9  \\ 
3FGL J0505.5+0416    &   3600    &  2016 Feb  05    &    1.1    &   16.9 &           3600 &     2016 Feb  05    &   1.1 & 16.9 \\
3FGL J0802.0+1005    &   2100    &  2016 Jan 28      &    1.3    &   18.5 &          2100  &    2016 Jan 28      &   1.3 & 18.5 \\ 
3FGL J0814.5+2943    &	3600    &	2016 Feb 06     &    1.3     &   18.3 &     	3600	 &	2016 Feb 06	&  1.3  &	18.3 	\\ 
3FGL J1107.5+0222 	   &	3000    &	2015 Dec 23     &    1.6     &   18.3 &    	3000	 &	2015 Dec 23	&  1.9  &	18.3 \\       
3FGL J1109.4+2411    &	3600    &	2016 Jan 28      &    1.0     &   17.8 &    	3600	 &	2016 Jan 21	&  0.9  &	18.0 	\\ 
3FGL J1450.9+5200    &   1500     &  2015 Mar 14     &    2.2      & 18.4  &        1800   &    2015 Mar 14     &  2.1 & 18.4 \\
3FGL J2116.1+3339	   &	  450    &	2015 Dec 24     &   1.9       &  16.7 &  	450	 &	2015 Dec 24	&  1.9  &	16.7	\\ 
\hline
\end{tabular}
\tablenotetext{}{
\raggedright
\footnotesize \texttt{Col.1}: Name of the target; \texttt{Col.2}: Total integration time with the Grism B; \texttt{Col.3}: Date of Observation with Grism B; \texttt{Col.4}: Seeing during the observation with the Grism B; \texttt{Col.5}: r ABmag measured on the acquisition image for the Grism B; \texttt{Col.6}: Total integration time with the Grism R; \texttt{Col.7}: Date of Observation with Grism R; \texttt{Col.8}: Seeing during the observation with the Grism R; \texttt{Col.9}: r ABmag measured on the acquisition images for the Grism R.}
\end{table*}

\newpage

\begin{table*}
\caption{PROPERTIES OF THE OPTICAL SPECTRA OF 3FGL/LAC SOURCES}\label{tab:table3}
\centering
\begin{tabular}{ l crcl}
\hline
OBJECT                       & $\alpha$  &   SNR          &   EW$_{min}$   &  z                                   \\
\hline
3FGL J0008.0+4713    &  +0.74*    &    10 - 50       &    0.50 - 3.10       &    $>$~1.659$^{i}$       \\ 
3FGL J0049.7+0237    &   -0.01      &    27 - 95       &    0.30 - 1.20      &    $>$~0.55$^{ll}$         \\ 
3FGL J0243.5+7119    &   -0.95     &    10 - 65        &    0.45 - 3.25      &    $>$~0.45$^{ll}$         \\ 
3FGL J0505.5+0416    &  -0.62      &    40 - 150      &    0.15 - 0.65      &    0.423$^{g}$               \\
3FGL J0802.0+1005    &  -0.77      &    32 - 108      &    0.30 - 0.70      &     $>$~0.58$^{ll}$         \\ 
3FGL J0814.5+2943    &  -0.95      &    54 - 161      &    0.20 - 0.55      &     0.703$^{g}$               \\ 
3FGL J1107.5+0222	   &  -0.82      &    27 - 208      &    0.15 - 1.15      &     $>$~1.0735$^{i}$      \\   
3FGL J1109.4+2411     &  -0.50     &    22 - 71        &    0.35 - 1.15       &     $>$~0.50$^{ll}$         \\ 
3FGL J1450.9+5200    &  -0.73      &    32 - 119      &    0.25 - 1.00       &     $>$~2.470$^{i}$      \\  
3FGL J2116.1+3339	   &  -1.66      &    40 - 130      &    0.30 - 0.85       &     $>$~0.25$^{ll}$         \\ 
\hline
\end{tabular}
\tablenotetext{}{
\raggedright
\footnotesize \texttt{Col.1}: Name of the target; \texttt{Col.2}: Optical spectral index derived from a power law fit in the range 4250-10000; \texttt{Col.3}: Range of S/N of the spectrum; \texttt{Col.4}: Range of the minimum equivalenth width (EW$_{min}$) derived from different regions of the spectrum (see text), \texttt{Col.5}: Proposed redshift from this work. The superscript letters indicate \textit{g} = host galaxy absorption, \textit{i}= intervening absorption, \textit{ll} = lower limit  of the redshift by assuming a BL Lac host galaxy with $M_{R}$ = -22.9}
\end{table*}

\begin{table*}
\caption{MEASUREMENTS OF SPECTRAL LINES}\label{tab:line}
\centering
\begin{tabular}{lccll}
\hline
OBJECT            &  $\lambda_{obs}$    &    EW (observed)    &     Line ID    &   z$_{line}$   \\
                  &  $\textrm{\AA}$             &     $\textrm{\AA}$           &                &                                                 \\
\hline
3FGL J0008.0+4713       &   7434.94    &   5.3         &      Mg II  (2796)      &  1.659 \\ 
                                        &   7454.20     &   4.2       &      Mg II  (2803)      &  1.659 \\ 
3FGL J0505.5+0416       &   5598.08      &  0.8        &      Ca II (3934)       &   0.423 \\
                                        &   5646.46     &  0.6         &      Ca II (3968)       &   0.423 \\
                                        &   6126.02     &  0.7         &    G-band (4305)     &  0.423  \\
3FGL J0814.5+2943       &   6699.60    & 0.6       &    Ca II (3934)   & 0.703                   \\ 
                                        &   6757.50    & 0.5       &    Ca II (3968)   & 0.703                  \\ 
3FGL J1107.5+0222	      &    5797.50    &  2.0     &    Mg II  (2796)    &  1.0735  \\  
                                        &    5812.02    &  1.9    &    Mg II  (2803)    &  1.0735 \\  
3FGL J1450.9+5200       &  4219.69     &   5.8      &   Ly$_{\alpha}$ (1216) &  2.470      \\   
                                        &  5127.64    &  1.1      &   C IV (1548)  &   2.312   \\ 
                                        &  5372.25    &  3.2      &   C IV (1548)  &   2.470     \\      
\hline
\end{tabular}
\tablenotetext{}{
\raggedright
\footnotesize \texttt{Col.1}: Name of the target; \texttt{Col.2}: Barycenter of the detected line (the superscript letter $a$ indicate absorption line); \texttt{Col.3}: Measured equivalent width; \texttt{Col.4}: Line identification; \texttt{Col.5}: Spectroscopic redshift.}
\end{table*}

\newpage

\begin{figure*}
\includegraphics[width=0.4\textwidth, angle=-90]{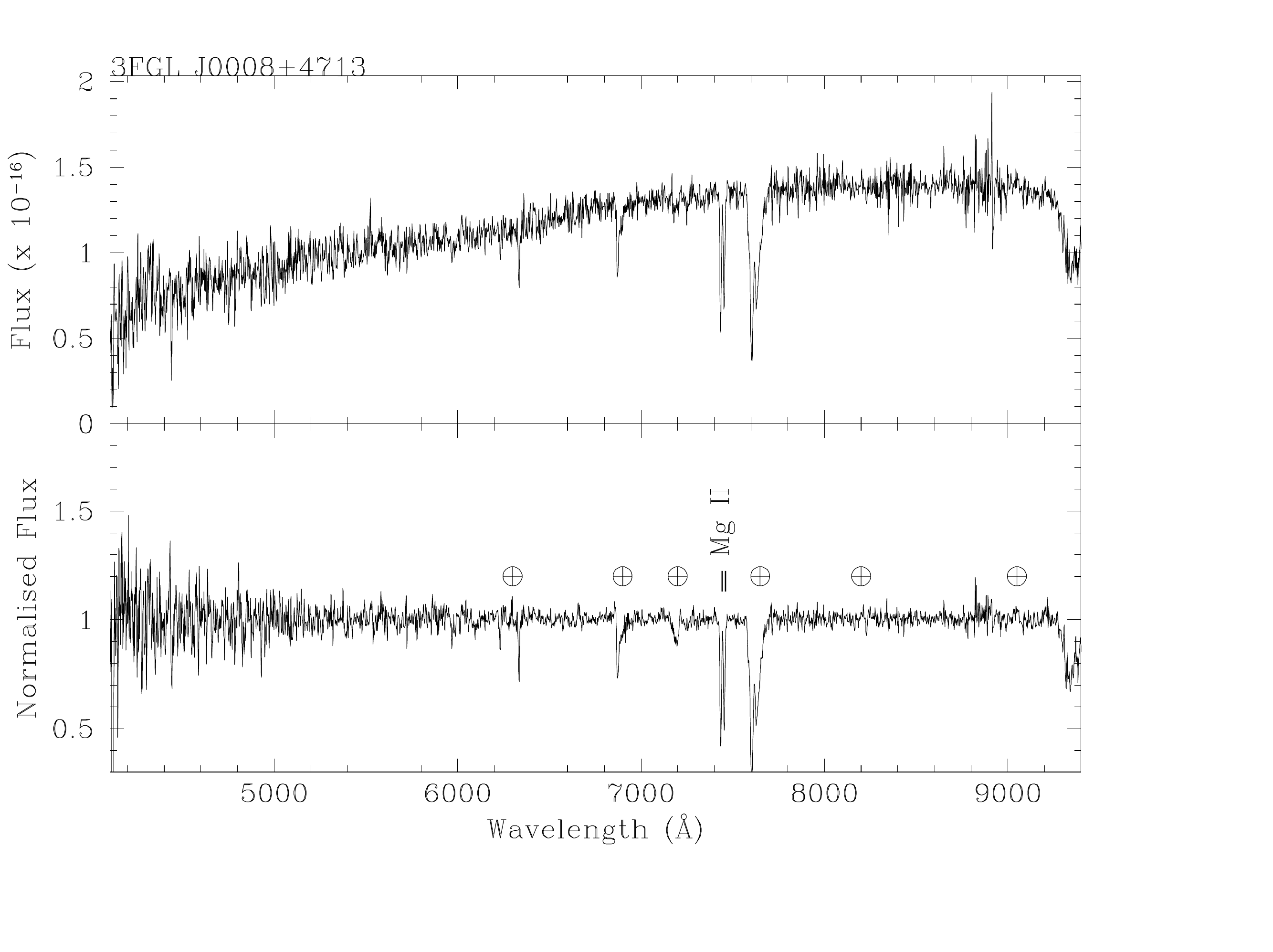}
\includegraphics[width=0.4\textwidth, angle=-90]{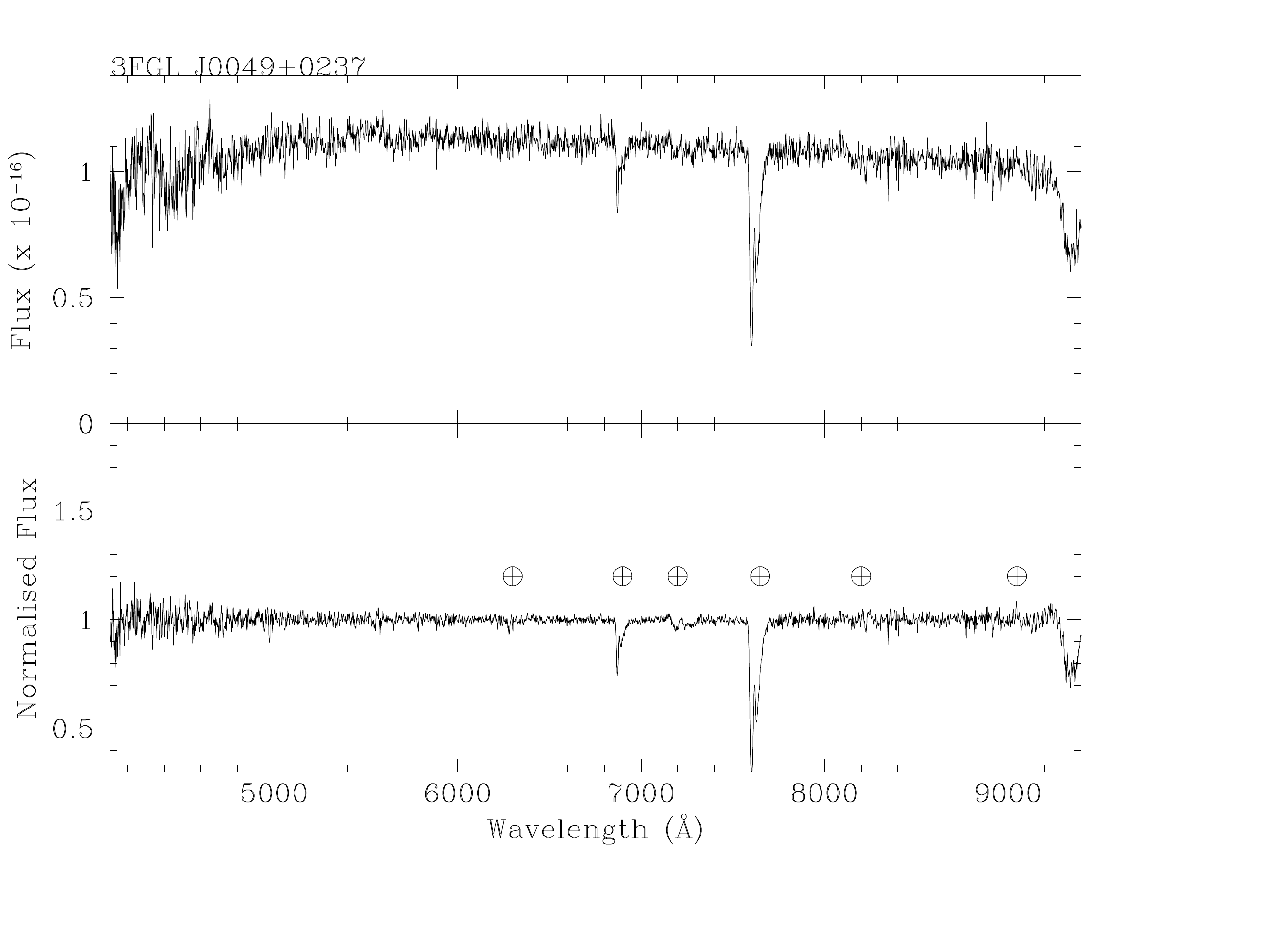}
\includegraphics[width=0.4\textwidth, angle=-90]{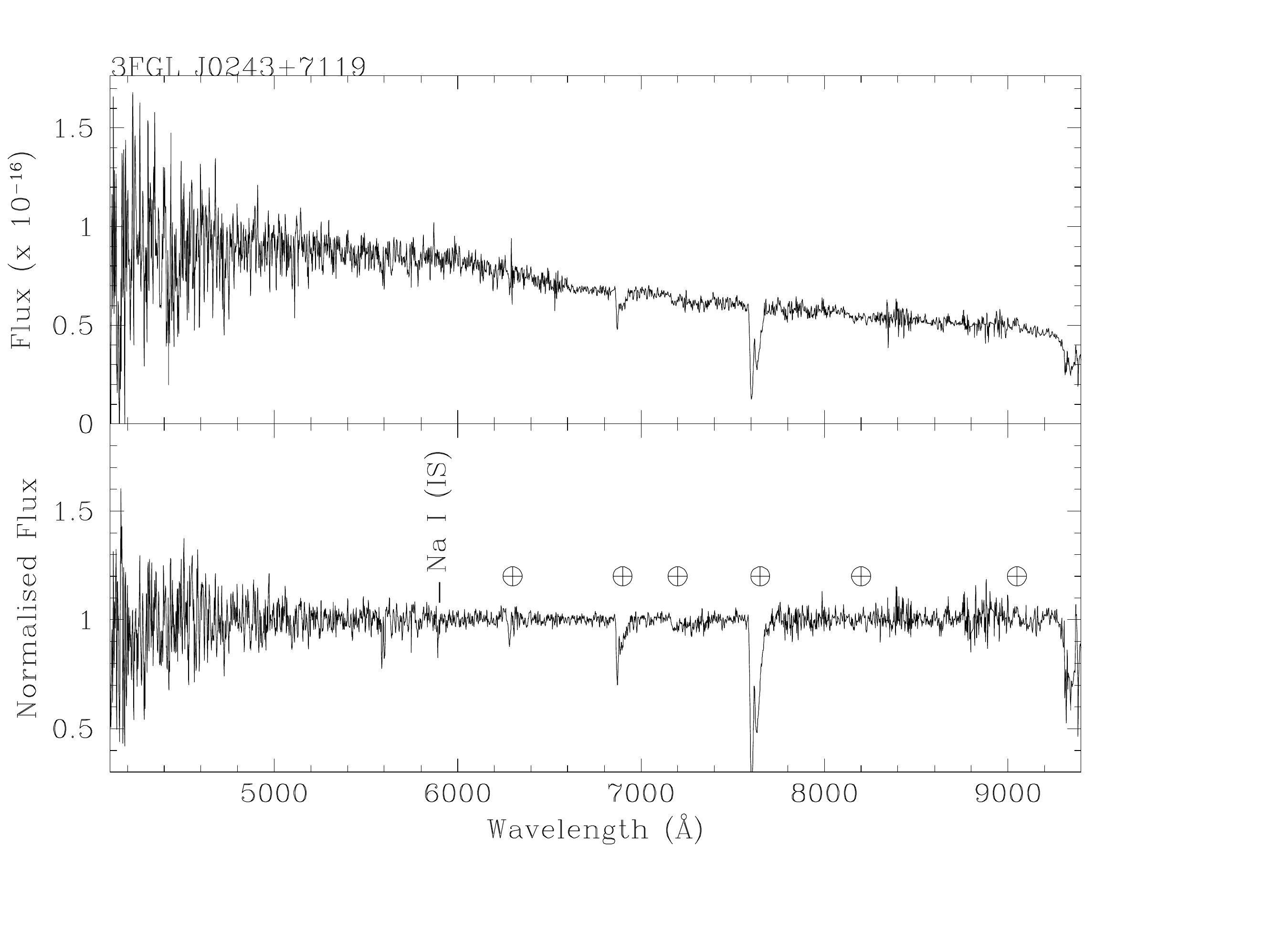} 
\includegraphics[width=0.4\textwidth, angle=-90]{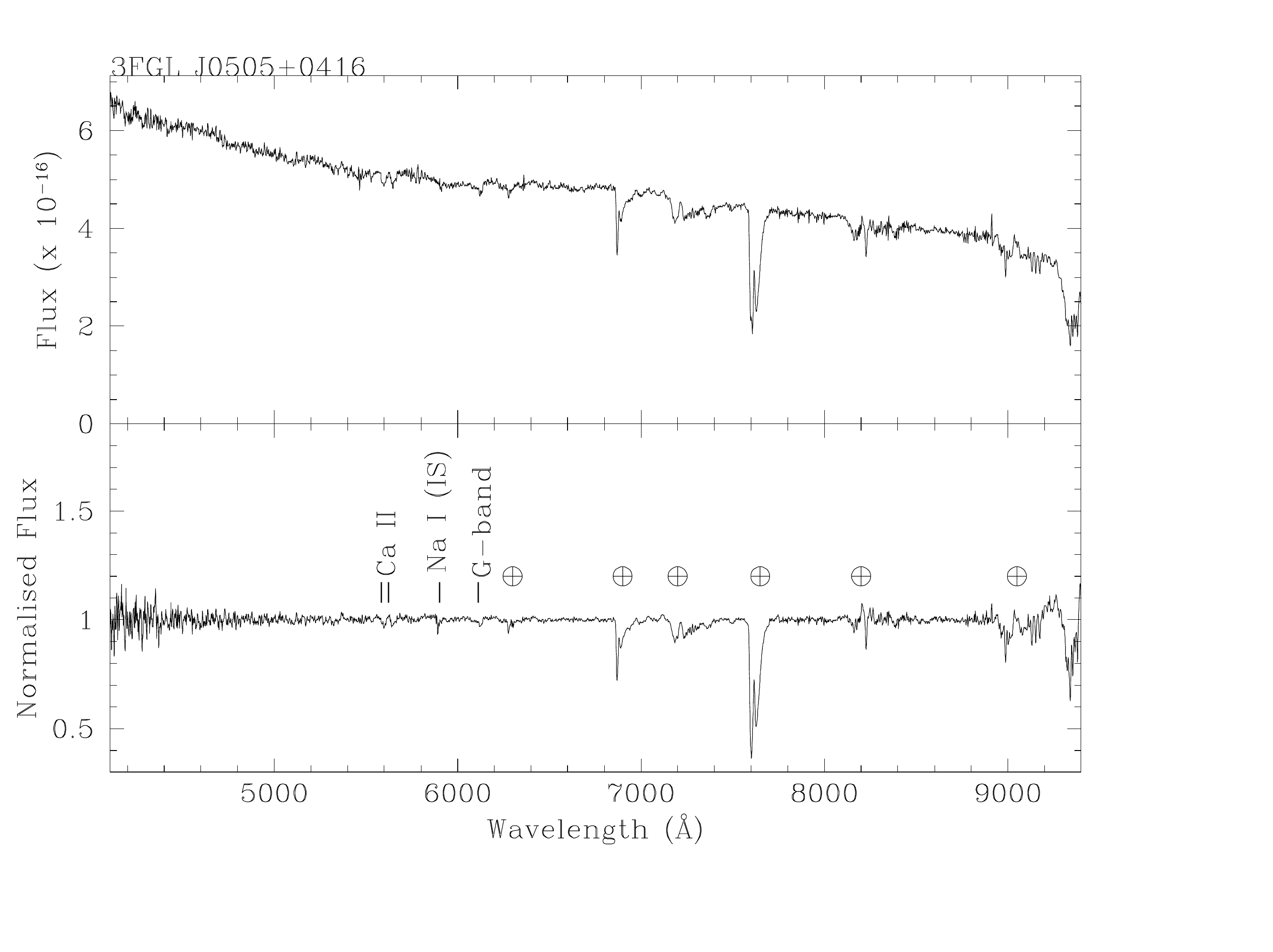}
\includegraphics[width=0.4\textwidth, angle=-90]{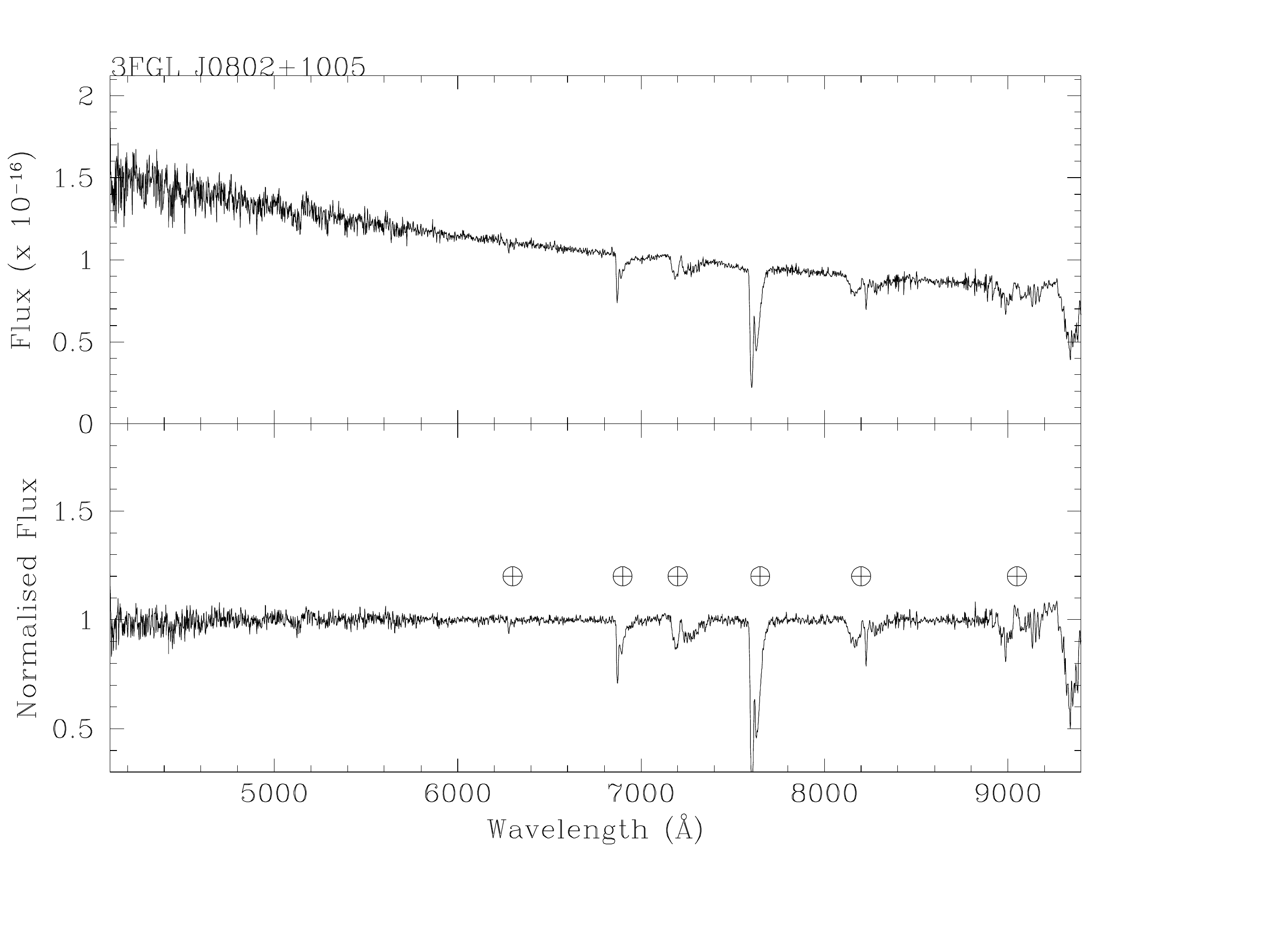}
\includegraphics[width=0.4\textwidth, angle=-90]{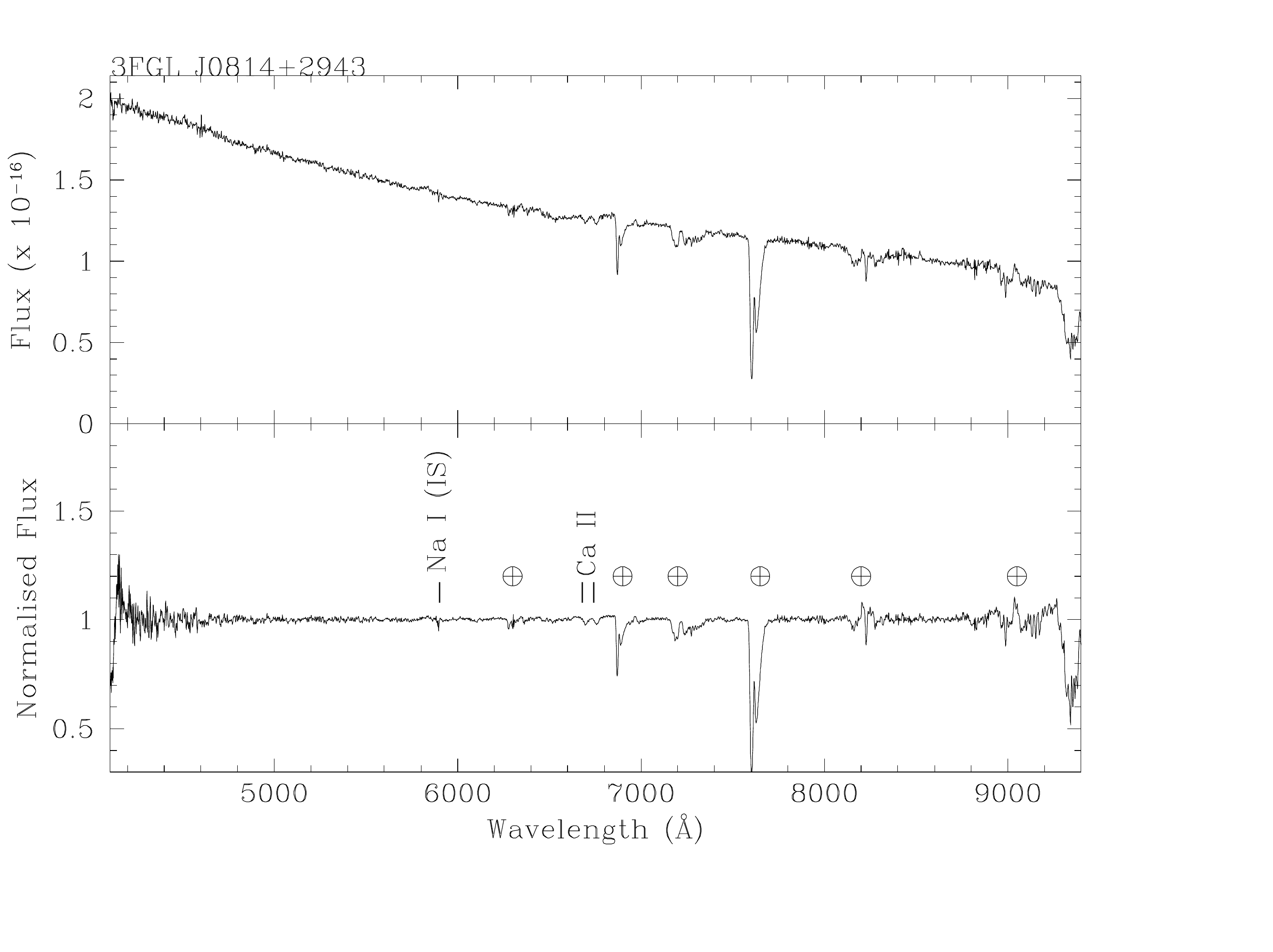}
\caption{Spectra of the high redshift 3FGL/LAT BLLs obtained at GTC. \textit{Top panel}: Flux calibrated and dereddered spectra. \textit{Bottom panel}: Normalized spectra. The main telluric bands are indicated by $\oplus$, the absorption features from interstellar medium of our galaxies are labelled as IS (Inter-Stellar). Data available at URL: http://www.oapd.inaf.it/zbllac/.}
\label{fig:spectra}
\end{figure*}

\setcounter{figure}{0}
\begin{figure*}
\includegraphics[width=0.4\textwidth, angle=-90]{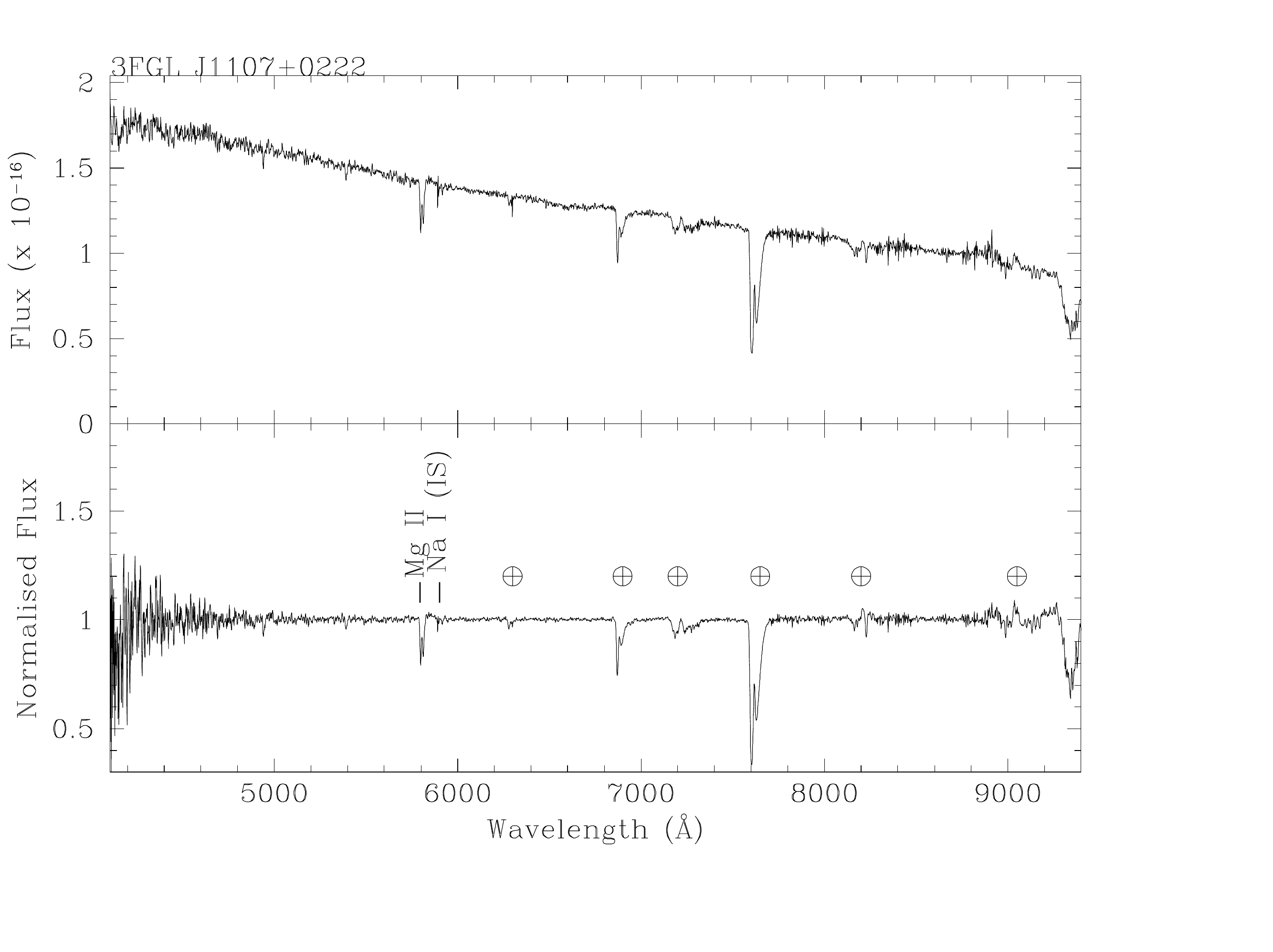}
\includegraphics[width=0.4\textwidth, angle=-90]{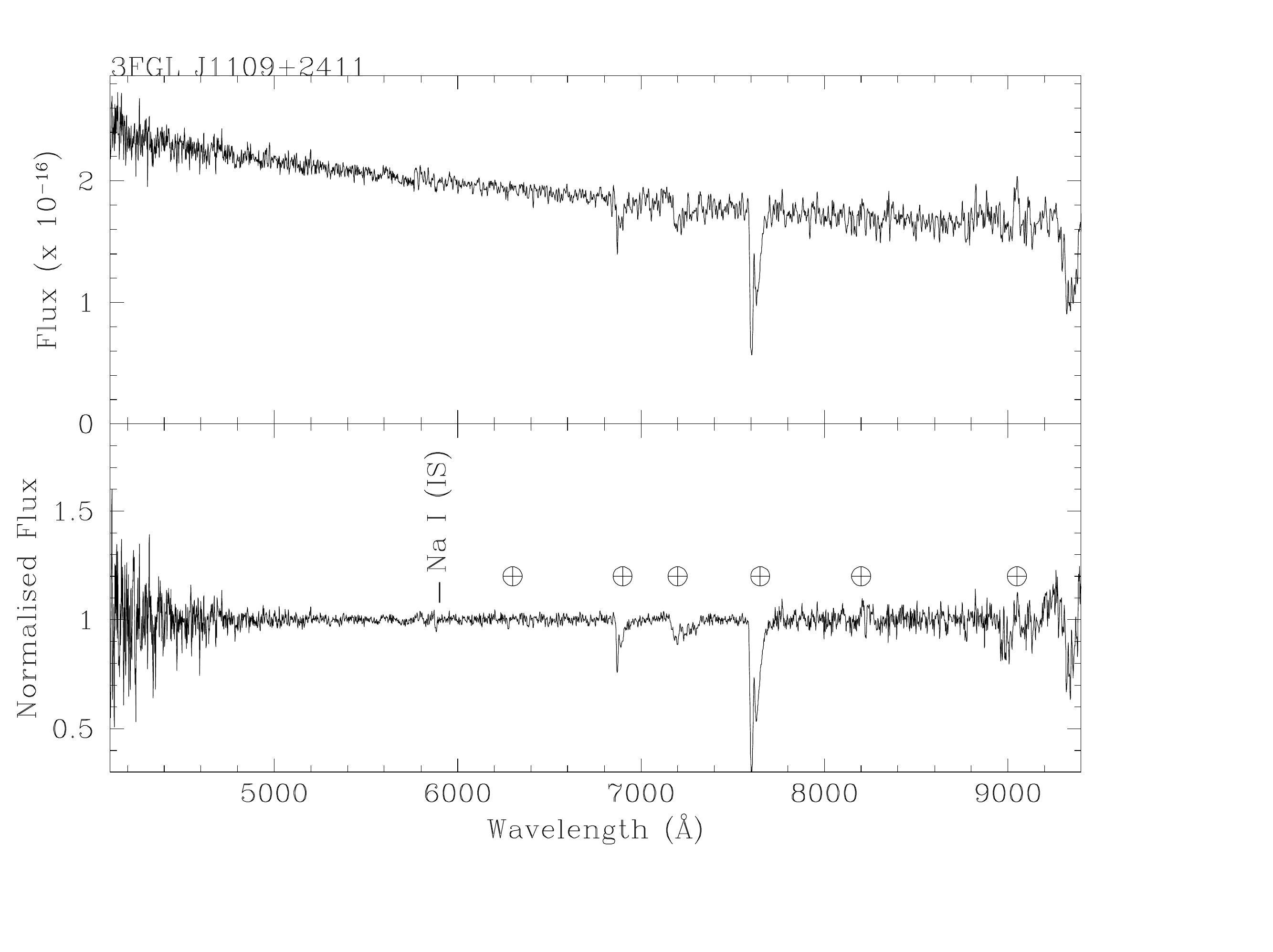} 
\includegraphics[width=0.4\textwidth, angle=-90]{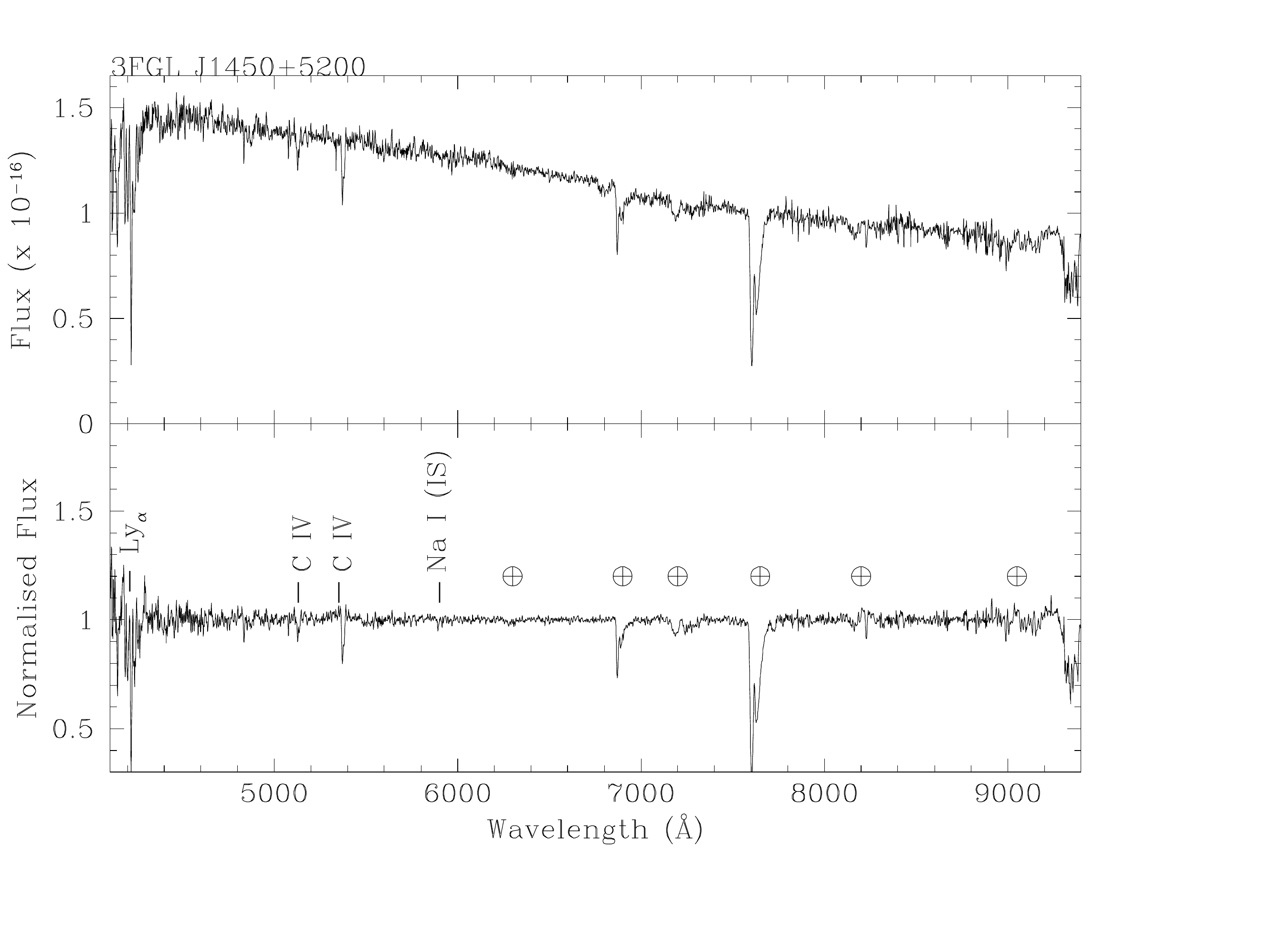} 
\includegraphics[width=0.4\textwidth, angle=-90]{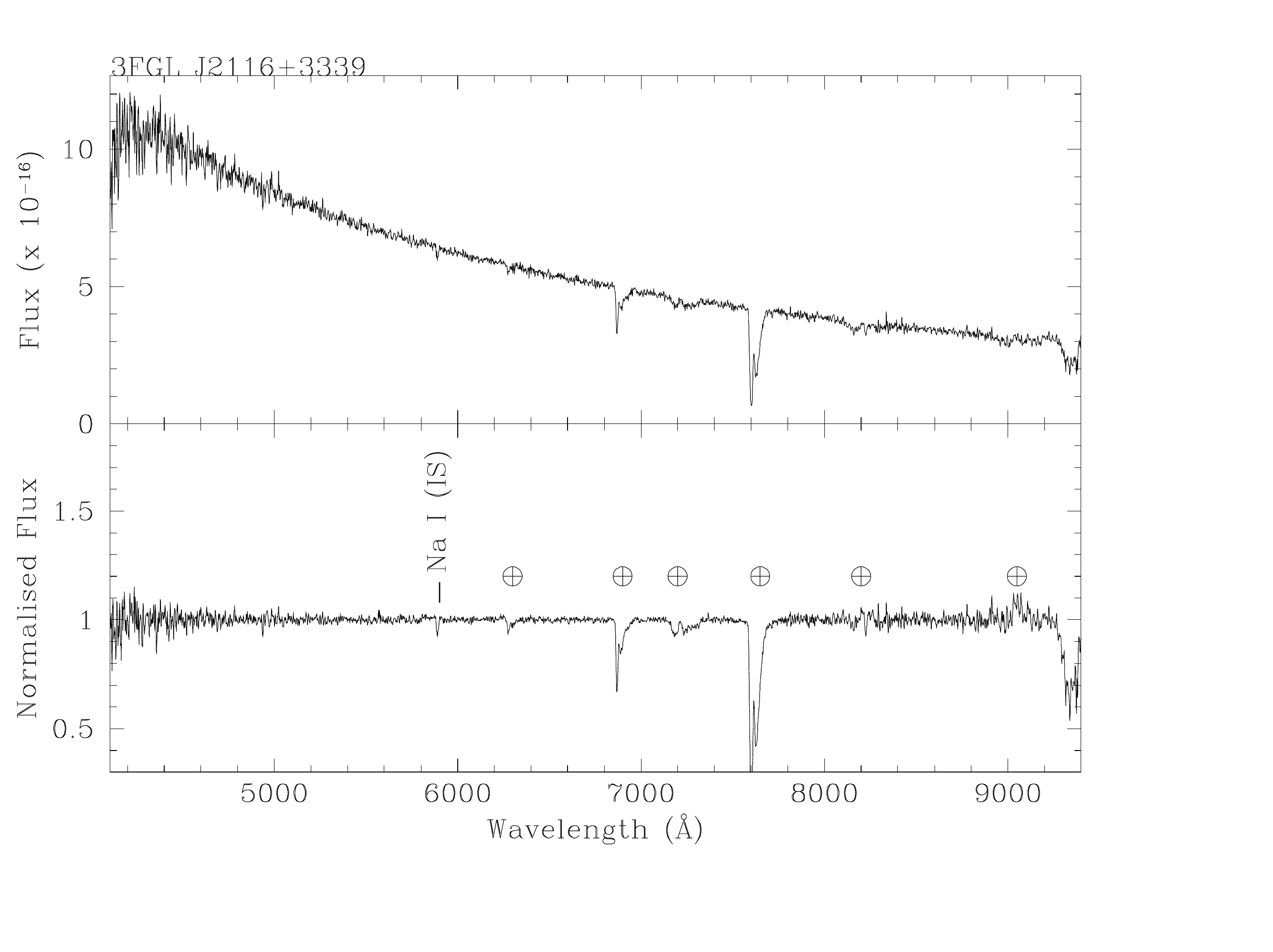} 
\caption{Continued. Data available at URL: http://www.oapd.inaf.it/zbllac/}
\end{figure*}

\newpage
\begin{figure*}
 \includegraphics[width=0.4\textwidth, angle=-90]{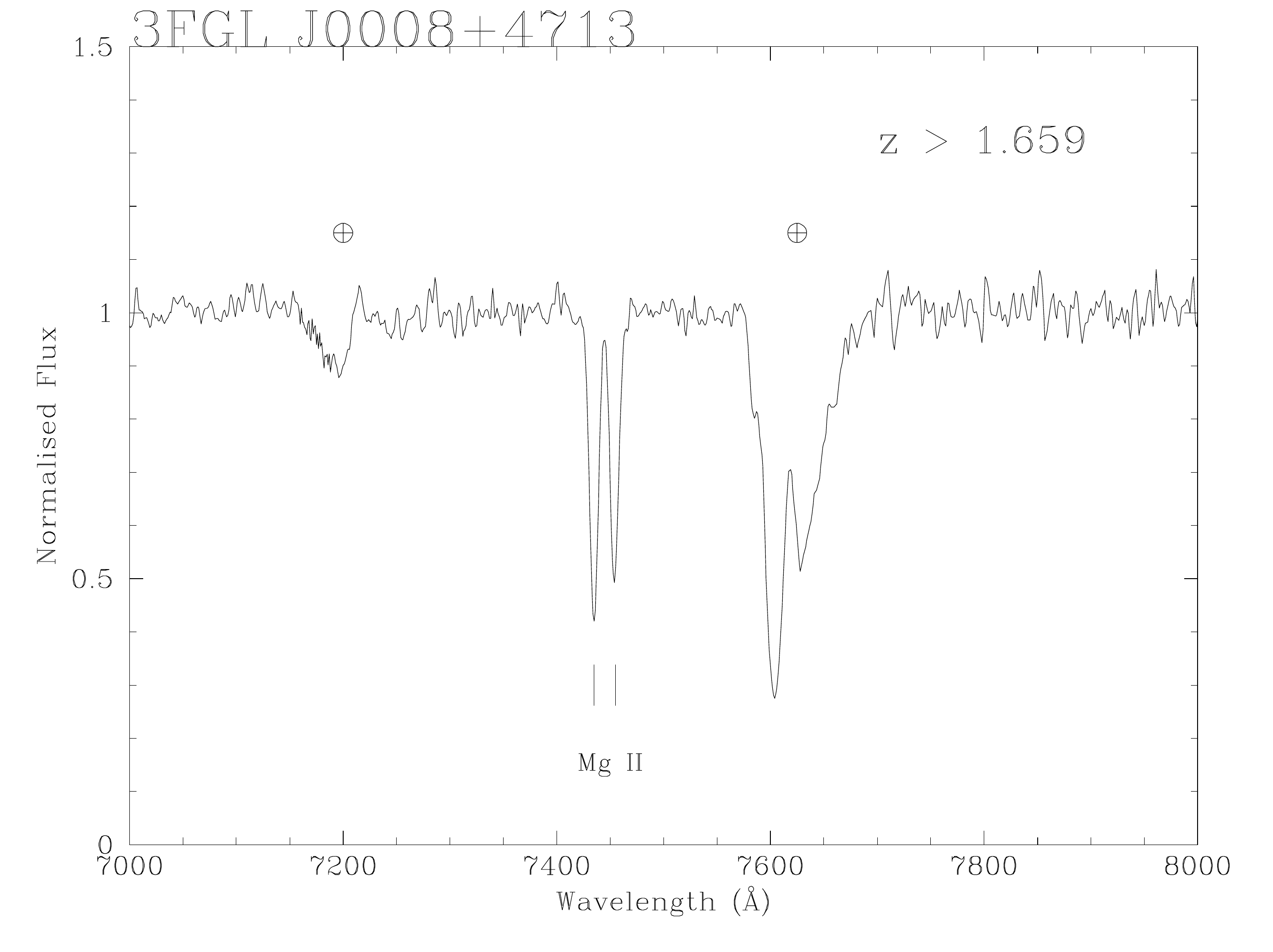}
 \includegraphics[width=0.4\textwidth, angle=-90]{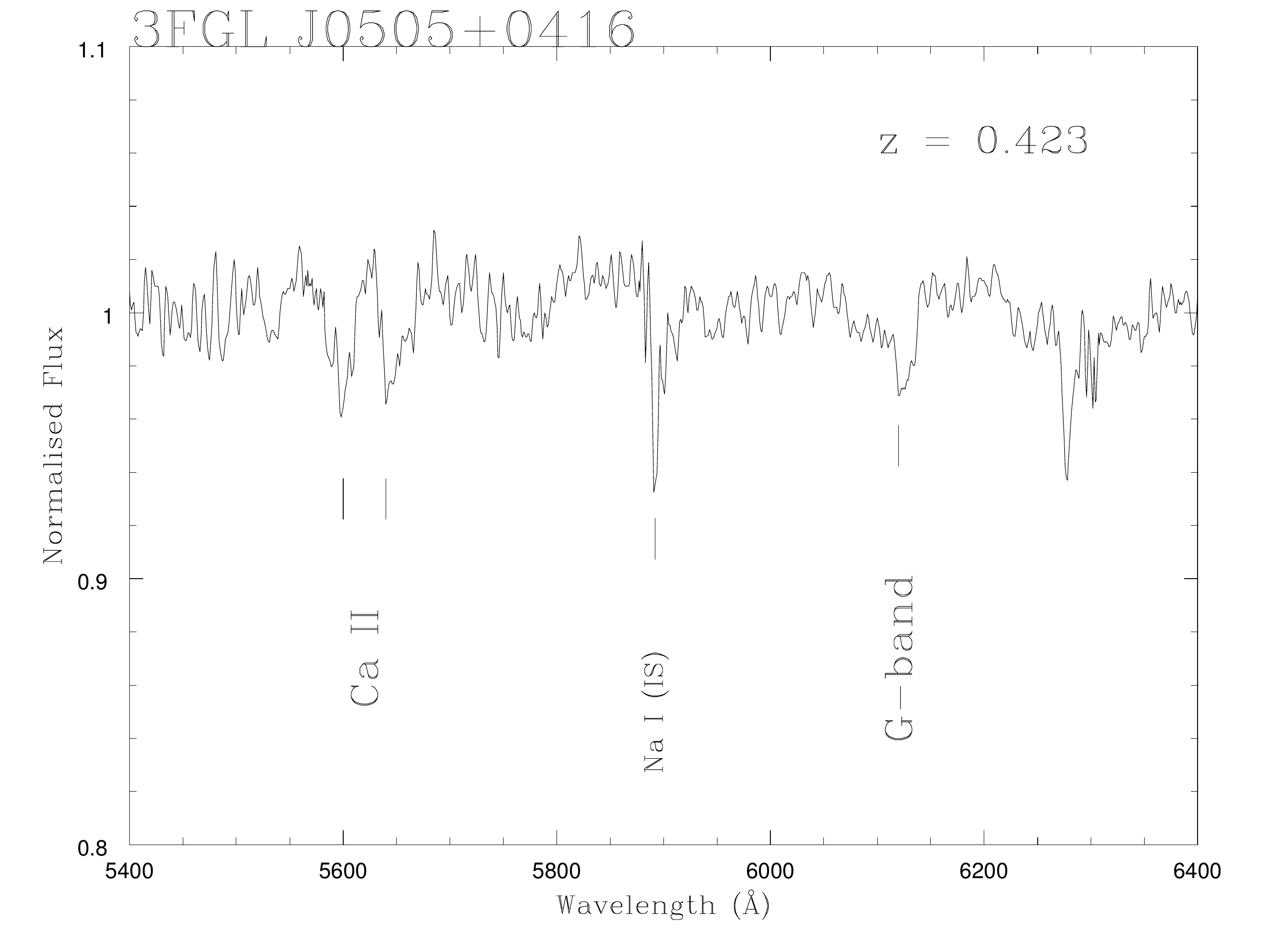}
  \includegraphics[width=0.4\textwidth, angle=-90]{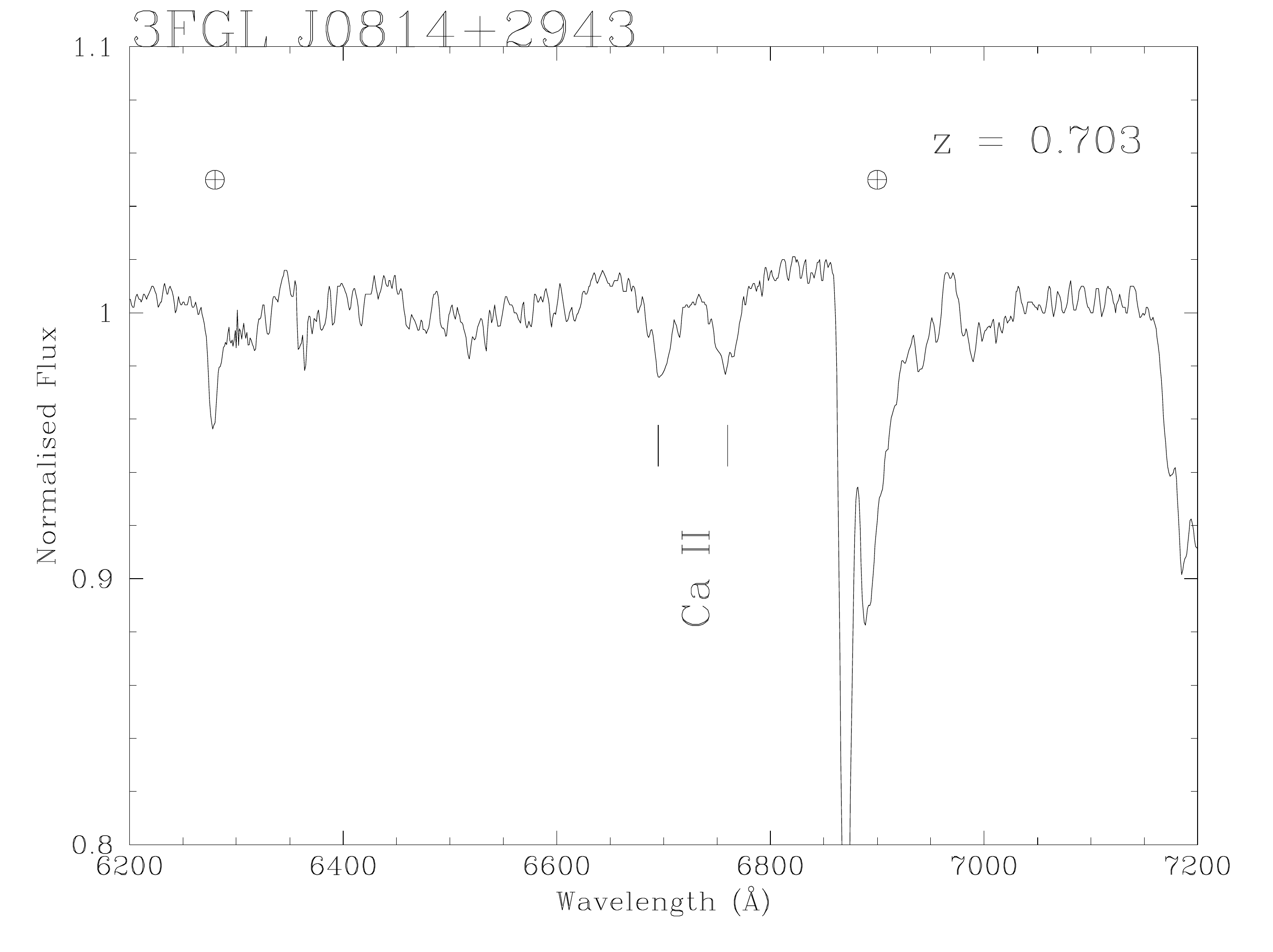} 
   \includegraphics[width=0.4\textwidth, angle=-90]{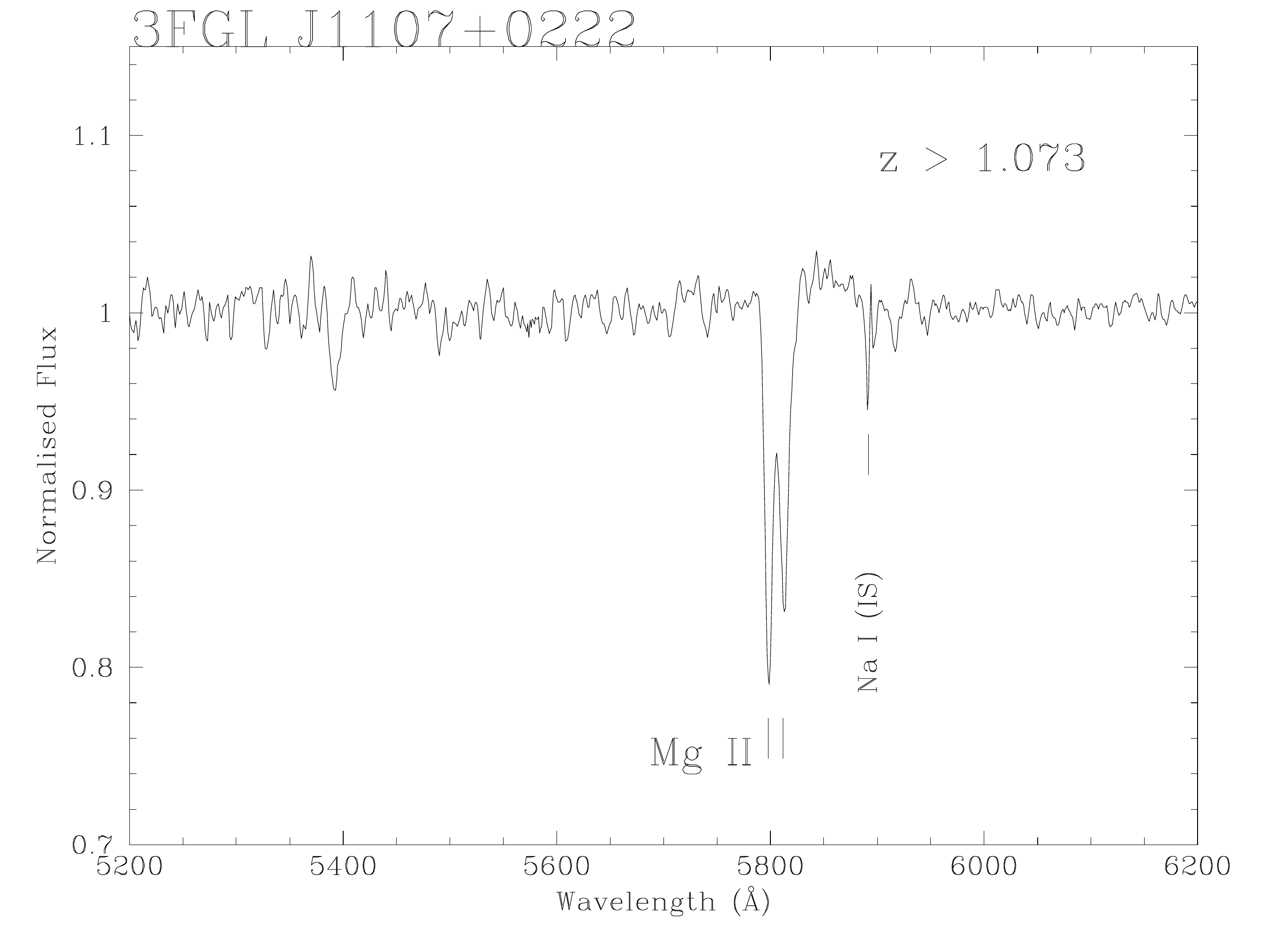}
   \includegraphics[width=0.4\textwidth, angle=-90]{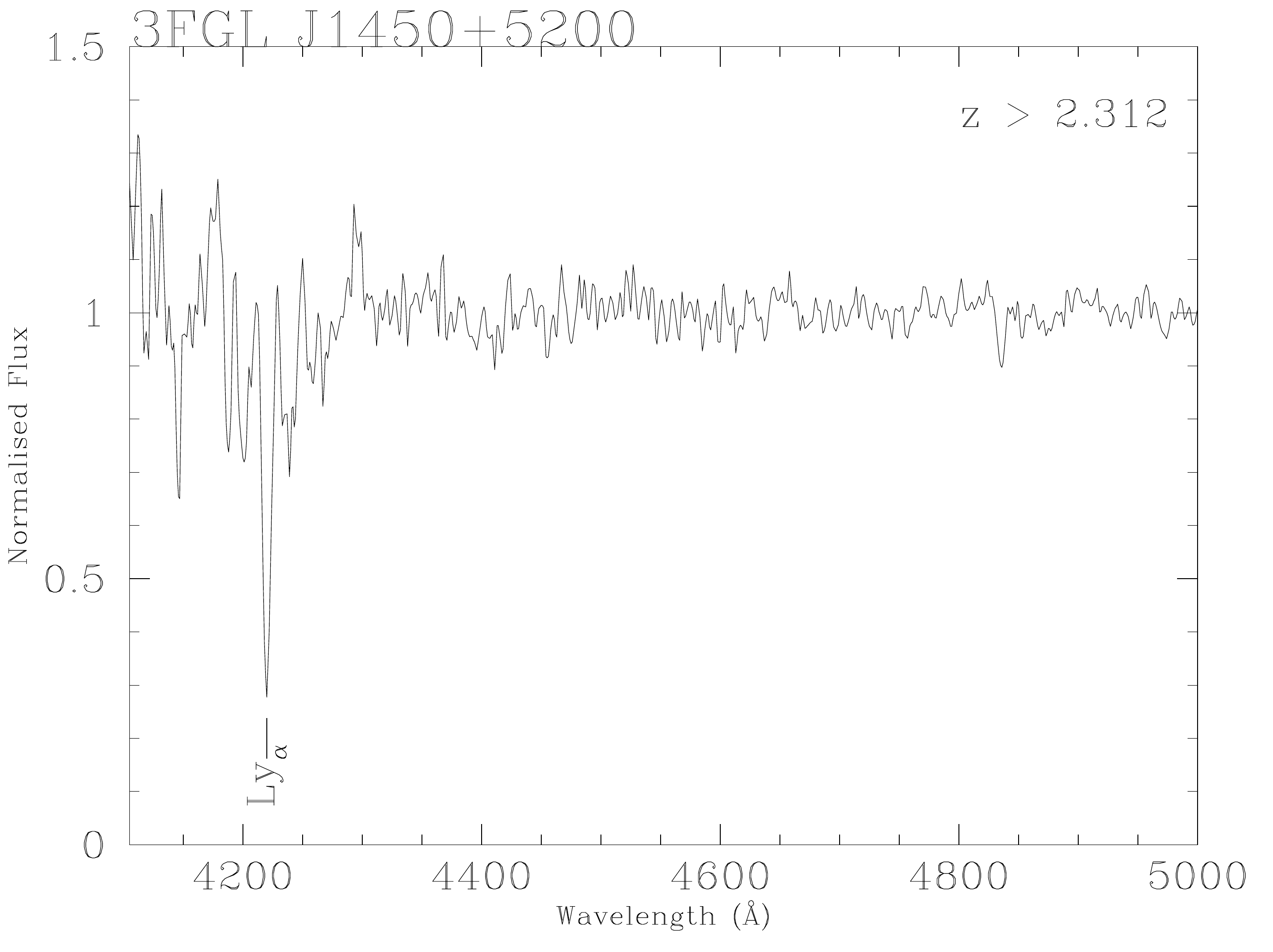}
   \includegraphics[width=0.4\textwidth, angle=-90]{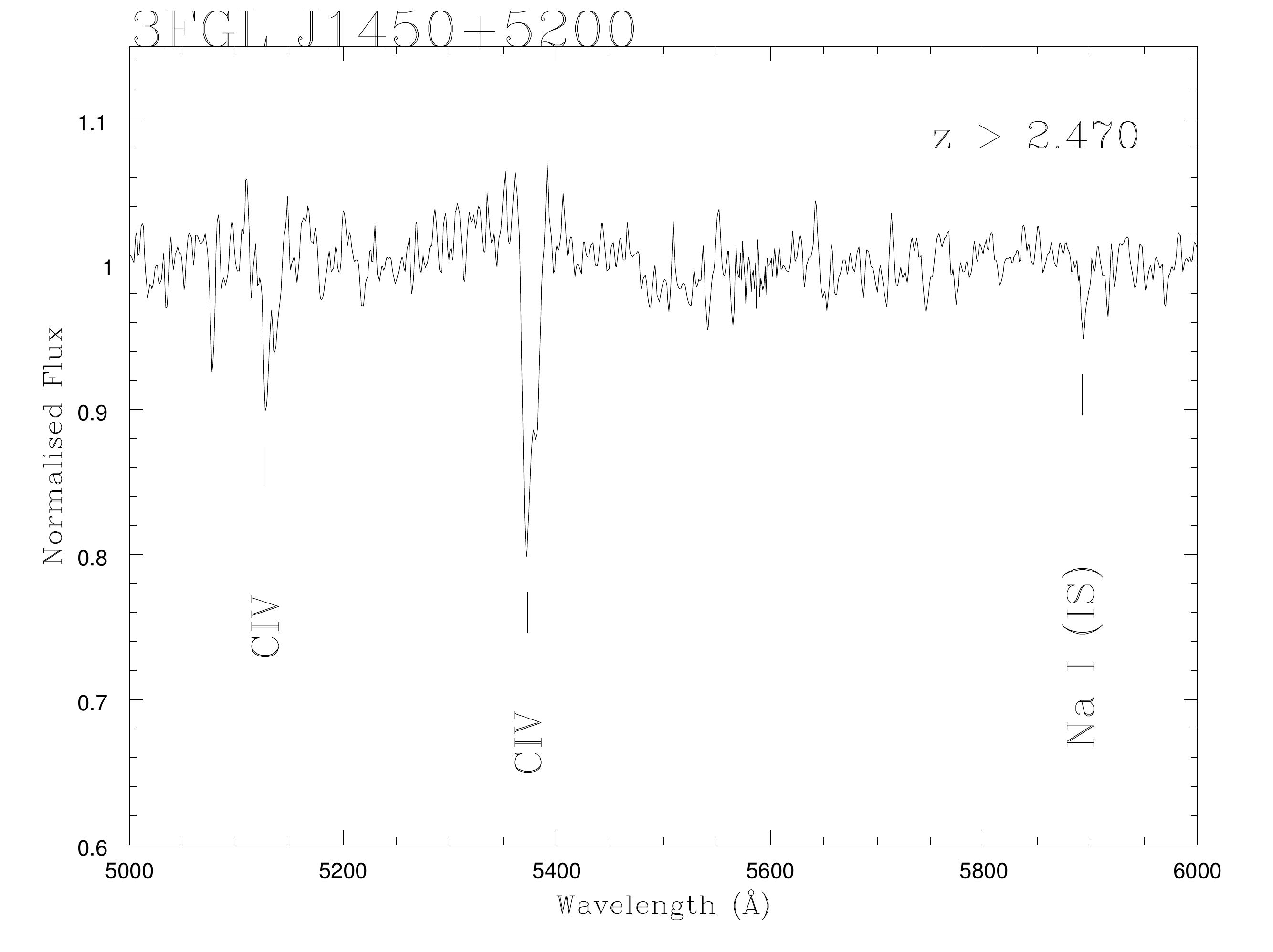}
 \caption{Close-up of the normalized spectra around the detected spectral features of the high redshift 3FGL/LAT BLLs obtained at GTC. Main telluric bands are indicated as $\oplus$, spectral lines are marked by line identification. } 
   \label{fig:spectraCU}
\end{figure*}


\bibliographystyle{aasjournal}
\bibliography{3FGL_GTC_biblio}

\begin{thebibliography}{}
\expandafter\ifx\csname natexlab\endcsname\relax\def\natexlab#1{#1}\fi

\bibitem[{{Acero} \& {Fermi-LAT Collaboration}(2015)}]{acero2015}
{Acero}, F., \& {Fermi-LAT Collaboration}. 2015, \apjs, 218, 23

\bibitem[{{Ackermann} {et~al.}(2015){Ackermann}, {Ajello}, {Atwood}, {Baldini},
  {Ballet}, {Barbiellini}, \& et~al.}]{3lac}
{Ackermann}, M., {Ajello}, M., {Atwood}, W.~B., {et~al.} 2015, \apj, 810, 14

\bibitem[{{Ackermann} {et~al.}(2017){Ackermann}, {Ajello}, {Baldini}, {Ballet},
  \& {Barbiellini}}]{ackermann2017}
{Ackermann}, M., {Ajello}, M., {Baldini}, L., {Ballet}, J., \& {Barbiellini},
  G. 2017, \apjl, 837, L5

\bibitem[{{Allington-Smith} {et~al.}(1991){Allington-Smith}, {Peacock}, \&
  {Dunlop}}]{allington1991}
{Allington-Smith}, J.~R., {Peacock}, J.~A., \& {Dunlop}, J.~S. 1991, \mnras,
  253, 287

\bibitem[{{Atwood} {et~al.}(2009){Atwood}, {Abdo}, {Ackermann}, {Althouse},
  {Anderson}, {Axelsson}, {Baldini}, {Ballet}, {Band}, \&
  {Barbiellini}}]{atwood2009}
{Atwood}, W.~B., {Abdo}, A.~A., {Ackermann}, M., {et~al.} 2009, \apj, 697, 1071

\bibitem[{{Bauer} {et~al.}(2000){Bauer}, {Condon}, {Thuan}, \&
  {Broderick}}]{bauer2000}
{Bauer}, F.~E., {Condon}, J.~J., {Thuan}, T.~X., \& {Broderick}, J.~J. 2000,
  \apjs, 129, 547

\bibitem[{{Bennett} {et~al.}(1986){Bennett}, {Lawrence}, {Burke}, {Hewitt}, \&
  {Mahoney}}]{bennett1986}
{Bennett}, C.~L., {Lawrence}, C.~R., {Burke}, B.~F., {Hewitt}, J.~N., \&
  {Mahoney}, J. 1986, \apjs, 61, 1

\bibitem[{{Cepa} {et~al.}(2003){Cepa}, {Aguiar-Gonzalez}, {Bland-Hawthorn},
  {Castaneda}, {Cobos}, {Correa}, {Espejo}, {Fragoso-Lopez}, {Fuentes},
  {Gigante}, {Gonzalez}, {Gonzalez-Escalera}, {Gonzalez-Serrano},
  {Joven-Alvarez}, {Lopez-Ruiz}, {Militello}, {Cano}, {Perez}, {Perez},
  {Rasilla}, {Sanchez}, \& {Tejada}}]{cepa2003}
{Cepa}, J., {Aguiar-Gonzalez}, M., {Bland-Hawthorn}, J., {et~al.} 2003, in
  Proc. SPIE, Vol. 4841, -, 1739--1749

\bibitem[{{Condon} {et~al.}(1998){Condon}, {Cotton}, {Greisen}, {Yin},
  {Perley}, {Taylor}, \& {Broderick}}]{nvss1998}
{Condon}, J.~J., {Cotton}, W.~D., {Greisen}, E.~W., {et~al.} 1998, \aj, 115,
  1693

\bibitem[{{Dermer} \& {Schlickeiser}(1993)}]{dermer1993}
{Dermer}, C.~D., \& {Schlickeiser}, R. 1993, \apj, 416, 458

\bibitem[{{Dunlop} {et~al.}(1989){Dunlop}, {Peacock}, {Savage}, {Lilly},
  {Heasley}, \& {Simon}}]{dunlop1989}
{Dunlop}, J.~S., {Peacock}, J.~A., {Savage}, A., {et~al.} 1989, \mnras, 238,
  1171

\bibitem[{{Falomo} \& {Kotilainen}(1999)}]{falomo1999}
{Falomo}, R., \& {Kotilainen}, J.~K. 1999, \aap, 352, 85

\bibitem[{{Falomo} {et~al.}(2014){Falomo}, {Pian}, \& {Treves}}]{falomo2014}
{Falomo}, R., {Pian}, E., \& {Treves}, A. 2014, \aapr, 22, 73

\bibitem[{{Fossati} {et~al.}(1998){Fossati}, {Maraschi}, {Celotti}, {Comastri},
  \& {Ghisellini}}]{fossati1998}
{Fossati}, G., {Maraschi}, L., {Celotti}, A., {Comastri}, A., \& {Ghisellini},
  G. 1998, \mnras, 299, 433

\bibitem[{{Franceschini} \& {Rodighiero}(2017)}]{franceschini2017}
{Franceschini}, A., \& {Rodighiero}, G. 2017, \aap, in press

\bibitem[{{Franceschini} {et~al.}(2008){Franceschini}, {Rodighiero}, \&
  {Vaccari}}]{franceschini2008}
{Franceschini}, A., {Rodighiero}, G., \& {Vaccari}, M. 2008, \aap, 487, 837

\bibitem[{{Ghisellini} {et~al.}(2017){Ghisellini}, {Righi}, {Costamante}, \&
  {Tavecchio}}]{ghisellini2017}
{Ghisellini}, G., {Righi}, C., {Costamante}, L., \& {Tavecchio}, F. 2017, ArXiv
  e-prints, arXiv:1702.02571

\bibitem[{{Ghisellini} \& {Tavecchio}(2009)}]{ghisellini2009a}
{Ghisellini}, G., \& {Tavecchio}, F. 2009, \mnras, 397, 985

\bibitem[{{Kock} {et~al.}(1996){Kock}, {Meisenheimer}, {Brinkmann}, {Neumann},
  \& {Siebert}}]{kock1996}
{Kock}, A., {Meisenheimer}, K., {Brinkmann}, W., {Neumann}, M., \& {Siebert},
  J. 1996, \aap, 307, 745

\bibitem[{{Landoni} {et~al.}(2014){Landoni}, {Falomo}, {Treves}, \&
  {Sbarufatti}}]{landoni2014}
{Landoni}, M., {Falomo}, R., {Treves}, A., \& {Sbarufatti}, B. 2014, \aap, 570,
  A126

\bibitem[{{Landoni} {et~al.}(2013){Landoni}, {Falomo}, {Treves}, {Sbarufatti},
  {Barattini}, {Decarli}, \& {Kotilainen}}]{landoni2013}
{Landoni}, M., {Falomo}, R., {Treves}, A., {et~al.} 2013, \aj, 145, 114

\bibitem[{{Landoni} {et~al.}(2012){Landoni}, {Falomo}, {Treves}, {Sbarufatti},
  {Decarli}, {Tavecchio}, \& {Kotilainen}}]{landoni2012}
---. 2012, \aap, 543, A116

\bibitem[{{Landoni} {et~al.}(2015){Landoni}, {Falomo}, {Treves}, {Scarpa}, \&
  {Reverte Pay{\'a}}}]{landoni2015}
{Landoni}, M., {Falomo}, R., {Treves}, A., {Scarpa}, R., \& {Reverte Pay{\'a}},
  D. 2015, \aj, 150, 181

\bibitem[{{Laurent-Muehleisen} {et~al.}(1998){Laurent-Muehleisen}, {Kollgaard},
  {Ciardullo}, {Feigelson}, {Brinkmann}, \& {Siebert}}]{laurent1998}
{Laurent-Muehleisen}, S.~A., {Kollgaard}, R.~I., {Ciardullo}, R., {et~al.}
  1998, \apjs, 118, 127

\bibitem[{{Madejski} \& {Sikora}(2016)}]{madejski2016}
{Madejski}, G.~., \& {Sikora}, M. 2016, \araa, 54, 725

\bibitem[{{Maraschi} {et~al.}(1992){Maraschi}, {Ghisellini}, \&
  {Celotti}}]{maraschi1992}
{Maraschi}, L., {Ghisellini}, G., \& {Celotti}, A. 1992, \apjl, 397, L5

\bibitem[{{Massaro} {et~al.}(2015){Massaro}, {Maselli}, {Leto}, {Marchegiani},
  {Perri}, {Giommi}, \& {Piranomonte}}]{massaro2015}
{Massaro}, E., {Maselli}, A., {Leto}, C., {et~al.} 2015, \apss, 357, 75

\bibitem[{{Nilsson} {et~al.}(2003){Nilsson}, {Pursimo}, {Heidt}, {Takalo},
  {Sillanp{\"a}{\"a}}, \& {Brinkmann}}]{nilsson2003}
{Nilsson}, K., {Pursimo}, T., {Heidt}, J., {et~al.} 2003, \aap, 400, 95

\bibitem[{{Paiano} {et~al.}(2016){Paiano}, {Landoni}, {Falomo}, {Scarpa}, \&
  {Treves}}]{paiano2016}
{Paiano}, S., {Landoni}, M., {Falomo}, R., {Scarpa}, R., \& {Treves}, A. 2016,
  \mnras, 458, 2836

\bibitem[{{Paiano} {et~al.}(2017){Paiano}, {Landoni}, {Falomo}, {Treves},
  {Scarpa}, \& {Righi}}]{paiano2017}
{Paiano}, S., {Landoni}, M., {Falomo}, R., {et~al.} 2017, \apj, 837, 144

\bibitem[{{Perlman} {et~al.}(1996){Perlman}, {Stocke}, {Schachter}, {Elvis},
  {Ellingson}, {Urry}, {Potter}, {Impey}, \& {Kolchinsky}}]{perlman1996}
{Perlman}, E.~S., {Stocke}, J.~T., {Schachter}, J.~F., {et~al.} 1996, \apjs,
  104, 251

\bibitem[{{Pita} {et~al.}(2014){Pita}, {Goldoni}, {Boisson}, {Lenain}, {Punch},
  {G{\'e}rard}, {Hammer}, {Kaper}, \& {Sol}}]{pita2014}
{Pita}, S., {Goldoni}, P., {Boisson}, C., {et~al.} 2014, \aap, 565, A12

\bibitem[{{Plotkin} {et~al.}(2010){Plotkin}, {Anderson}, {Brandt},
  {Diamond-Stanic}, {Fan}, {Hall}, {Kimball}, {Richmond}, {Schneider},
  {Shemmer}, {Voges}, {York}, {Bahcall}, {Snedden}, {Bizyaev}, {Brewington},
  {Malanushenko}, {Malanushenko}, {Oravetz}, {Pan}, \& {Simmons}}]{plotkin2010}
{Plotkin}, R.~M., {Anderson}, S.~F., {Brandt}, W.~N., {et~al.} 2010, \aj, 139,
  390

\bibitem[{{Sandrinelli} {et~al.}(2013){Sandrinelli}, {Treves}, {Falomo},
  {Farina}, {Foschini}, {Landoni}, \& {Sbarufatti}}]{sandrinelli2013}
{Sandrinelli}, A., {Treves}, A., {Falomo}, R., {et~al.} 2013, \aj, 146, 163

\bibitem[{{Sbarufatti} {et~al.}(2005){Sbarufatti}, {Treves}, \&
  {Falomo}}]{sbarufatti2005}
{Sbarufatti}, B., {Treves}, A., \& {Falomo}, R. 2005, \apj, 635, 173

\bibitem[{{Sbarufatti} {et~al.}(2006){Sbarufatti}, {Treves}, {Falomo}, {Heidt},
  {Kotilainen}, \& {Scarpa}}]{sbarufatti2006}
{Sbarufatti}, B., {Treves}, A., {Falomo}, R., {et~al.} 2006, \aj, 132, 1

\bibitem[{{Shaw} {et~al.}(2013){Shaw}, {Romani}, {Cotter}, {Healey},
  {Michelson}, {Readhead}, {Richards}, {Max-Moerbeck}, {King}, \&
  {Potter}}]{shaw2013}
{Shaw}, M.~S., {Romani}, R.~W., {Cotter}, G., {et~al.} 2013, \apj, 764, 135

\bibitem[{{Stephens} {et~al.}(2015){Stephens}, {Ballet}, {Burnett},
  {Cavazzuti}, {Digel}, \& {Fermi LAT Collaboration}}]{3FGL}
{Stephens}, T.~E., {Ballet}, J., {Burnett}, T., {et~al.} 2015, in American
  Astronomical Society Meeting Abstracts, Vol. 225, American Astronomical
  Society Meeting Abstracts, 422.02

\bibitem[{{Stickel} \& {Kuehr}(1996)}]{stickel1996}
{Stickel}, M., \& {Kuehr}, H. 1996, \aaps, 115, 1

\bibitem[{{The Fermi-LAT Collaboration}(2017)}]{3fhlcatalog}
{The Fermi-LAT Collaboration}. 2017, ArXiv e-prints, arXiv:1702.00664

\bibitem[{{Voges} {et~al.}(1999){Voges}, {Aschenbach}, {Boller},
  {Br{\"a}uninger}, {Briel}, {Burkert}, {Dennerl}, {Englhauser}, {Gruber},
  {Haberl}, {Hartner}, {Hasinger}, {K{\"u}rster}, {Pfeffermann}, {Pietsch},
  {Predehl}, {Rosso}, {Schmitt}, {Tr{\"u}mper}, \& {Zimmermann}}]{voges1999}
{Voges}, W., {Aschenbach}, B., {Boller}, T., {et~al.} 1999, \aap, 349, 389

\bibitem[{{White} {et~al.}(2000){White}, {Becker}, {Gregg},
  {Laurent-Muehleisen}, {Brotherton}, {Impey}, {Petry}, {Foltz}, {Chaffee},
  {Richards}, {Oegerle}, {Helfand}, {McMahon}, \& {Cabanela}}]{white2000}
{White}, R.~L., {Becker}, R.~H., {Gregg}, M.~D., {et~al.} 2000, \apjs, 126, 133

\end{thebibliography}






\end{document}